# Cobalt-based Co3Mo3N/Co4N/Co Metallic Heterostructure as a Highly Active Electrocatalyst for Alkaline Overall Water Splitting

*Yuanwu Liu†, Lirong Wang†, René Hübner, Johannes Kresse, Xiaoming Zhang\*, Marielle Deconinck, Yana Vaynzof, Inez M. Weidinger, Alexander Eychmüller\**

Y. Liu, J. Kresse, Prof. A. Eychmüller

Physical Chemistry, TU Dresden, Zellescher Weg 19, 01069 Dresden, Germany

*E-mail: alexander.eychmueller@tu-dresden.de

L. Wang, Prof. X. Zhang

School of Materials Science and Engineering, Hebei University of Technology, Tianjin 300130, China

*E-mail: zhangxiaoming87@hebut.edu.cn

Dr. R. Hübner

Institute of Ion Beam Physics and Materials Research, Helmholtz-Zentrum Dresden-Rossendorf e.V., Bautzner Landstrasse 400, 01328 Dresden, Germany

M. Deconinck, Prof. Y. Vaynzof

Chair for Emerging Electronic Technologies, TU Dresden, Nöthnitzer Str. 61, Dresden, 01187 Sachsen, Germany

Leibniz-Institute for Solid State and Materials Research Dresden, Helmholtzstraße 20, Dresden, 01069 Sachsen, Germany

Prof. I. Weidinger

Fakultät Chemie und Lebensmittelchemie, Technische Universität Dresden, Zellescher Weg 19, 01069 Dresden, Germany

† Y. Liu and L. Wang contribute equally to this work.

**Abstract:** Alkaline water electrolysis is considered a commercially viable option for large-scale hydrogen production. However, this process still faces challenges due to the high voltage (>1.65 V at 10 mA cm$^{-2}$) and its limited stability at higher current densities due to the inefficient electron transport kinetics. Herein, a novel cobalt based metallic heterostructure (Co$_3$Mo$_3$N/Co$_4$N/Co) is designed for application for water electrolysis. Operando Raman experiments reveal that the formation of Co$_3$Mo$_3$N/Co$_4$N heterointerface boosts the free water adsorption and dissociation, resulting in a surplus of protons available for subsequent hydrogen production. Furthermore, the altered electronic structure of Co$_3$Mo$_3$N/Co$_4$N heterointerface optimizes the $\Delta G_H$ of nitrogen atoms at the interface. This synergistic effect between interfacial nitrogen atoms and metal phase cobalt creates highly efficient hydrogen evolution reaction (HER) active sites, thereby enhancing the overall performance. Additionally, the heterostructure exhibits a rapid OH$^-$ adsorption rate, coupled with a strong adsorption strength, leading to improved oxygen evolution reaction (OER) performance. Crucially, the metallic heterojunction facilitates fast electron transport, expediting the aforementioned reaction steps and ultimately improving the overall efficiency of water splitting. The water electrolyzer with Co$_3$Mo$_3$N/Co$_4$N/Co as a catalyst exhibits outstanding performance, requiring an impressively low cell voltage of 1.58 V at 10 mA cm$^{-2}$ and maintaining approximately 100% retention over a remarkable 100 h duration at 200 mA cm$^{-2}$. This performance significantly exceeds that of the commercial Pt/C || RuO$_2$ electrolyzer.

## 1. Introduction

Water electrolysis is a promising technology to address the global energy crisis by providing a green and scalable strategy to convert renewable electricity into hydrogen fuel. Compared with acidic electrolyzers, alkaline electrolyzers are more popular in the industry due to their cheaper ion exchange membranes and faster oxygen evolution reaction (OER) kinetics[1,2]. However, this brings forth a new disadvantage: hydrogen evolution reaction (HER) kinetics using an alkaline electrolyte are two orders of magnitude lower than that using an acidic electrolyte[3]. Based on earlier work, the bottleneck of alkaline water electrolysis, especially alkaline HER, is largely related to the sluggish water dissociation step (Volmer and Heyrovsky steps)[4]. In the Volmer step, the HO-H bond is broken as it interacts with an electron, ultimately forming an adsorbed Hydrogen atom ($H_{ad}$). In the Heyrovsky step, the $H_{ad}$ will combine with another broken HO-H bond and electron to form $H_2$. To date, noble metal-based catalysts (NMC) are still regarded as the state-of-the-art electrocatalysts due to their high conductivity, strong water dissociation ability and optimal $d$ band center for adsorbing $H^+/OH^-$[5,6]. NMC, however, are expensive and unstable in long-term reaction processes, limiting their large-scale application[7]. From the perspective of commercialization, it is urgent to develop non-noble metal catalysts with high catalytic activity and excellent durability.

In general, an ideal bifunctional non-noble metal alkaline electrocatalyst should satisfy the following conditions: 1) sufficient number of active sites for fast adsorption for $H_2O/OH^-$, and the active sites have a strong $H_2O$ dissociation ability; 2) the adsorption of H at the active sites is neither too strong nor too weak, i.e. the Gibbs free energy for $H_{ad}$ ($\Delta G_H$) is close to zero[8]; 3) high conductivity that enables rapid electron transport from the interior of the catalyst to the catalytic reaction interface; 4) excellent mechanical stability and catalytic durability. Recently, Cobalt based compounds, especially a class of interstitial compounds with noble metal properties (Cobalt nitrides), have exhibited great potential for alkaline water splitting. Metallic $Co_4N$ - a typical member of the cobalt nitride family - has been reported to have better OER performance than benchmark $IrO_2$ because the Co atom in $Co_4N$ is prone to adsorb $OH^-$[9,10]. Nevertheless, the Co atoms in $Co_4N$ have a strong adsorption of H and requires high energy to dissociate water, resulting in poor HER activity. Introducing metallic Co as HER sites is an effective method to optimal $\Delta G_H$ because Co has a more suitable ability to adsorb and desorb hydrogen[11,12]. Additionally, coupling $Co_4N$ with the secondary phase, such as $CeO_2$[9], $CoO$[13], $Co_2P$[14], NiFe double layer hydroxides[15], is widely used to modify the water dissociation energy. However, the majority of secondary phase coupling with $Co_4N$ that has been reported is semiconductor-based, which results in a suppressed electron transport efficiency in electrocatalysts. In order to create high-performance electrocatalysts for alkaline water electrolysis, it is vital to continue developing $Co_4N$ based metallic heterostructures with excellent electrical conductivity.

The incorporation of molybdenum atoms into Co$_4$N can result in the formation of an alternative metal phase known as Co$_3$Mo$_3$N.[16, 17] Unlike pure Co$_4$N, Co$_3$Mo$_3$N exhibits an enhanced d-state density near the Fermi level. Additionally, the introduction of molybdenum atoms into Co$_4$N prompts electron transfer from Mo to Co atoms. This electron transfer optimizes the capacity of Co atoms to efficiently adsorb and desorb H atoms. Consequently, the unique electronic structure of Co$_3$Mo$_3$N establishes it as a superb electrocatalyst for HER. The synergy of Co$_3$Mo$_3$N and Co$_4$N may ensure effective electron transfer, simultaneously addressing the challenge of insufficient HER activity observed in pure Co$_4$N. Consequently, Co$_3$Mo$_3$N emerges as an ideal material for constructing Co$_4$N-based metallic heterostructures.

In this work, we design a bifunctional electrocatalyst of Co$_3$Mo$_3$N/Co$_4$N/Co composite with metallic properties, which perfectly meets the above requirements of alkaline water electrolysis. The formation of Co$_3$Mo$_3$N/Co$_4$N interface plays a crucial role in enhancing the water adsorption capacity and significantly reducing the energy barrier for water dissociation. Additionally, the interface optimization influences the electronic structure of N atoms, reducing its $\Delta G_H$ close to 0, and the introduction of metallic cobalt further optimizes the $\Delta G_H$ of the electrocatalyst. As a result, the Co$_3$Mo$_3$N/Co$_4$N/Co composite demonstrates outstanding performance in HER. Moreover, the upshift of the transition metal *d* band center contributes to the increased adsorption capacity of cobalt sites in the electrocatalyst, particularly for OH$^-$, favoring the OER. Furthermore, the presence of metallic conductivity in Co$_3$Mo$_3$N/Co$_4$N/Co facilitates efficient electron transfer and transport within the catalyst and at the catalyst/electrolyte interface. By integrating the aforementioned enhancements, the Co$_3$Mo$_3$N/Co$_4$N/Co demonstrates remarkable achievements with low overpotentials of 220 mV for the OER and 78 mV for the HER in a 1 M KOH electrolyte at 10 mA cm$^{-2}$. Moreover, the water electrolyzer operates efficiently with an exceptionally low cell voltage of only 1.58 V at 10 mA cm$^{-2}$ and exhibits outstanding durability even under demanding conditions of 200 mA cm$^{-2}$. This work presents innovative perspectives on enhancing the overall water splitting performance of transition metal-based metallic heterostructures, with a focus on improving their stability under high current densities.

## 2. Results and discussion

### 2.1. Bifunctional electrocatalysts design and physical characterization

The synthesis of the metallic Co$_3$Mo$_3$N/Co$_4$N/Co heterostructure involves two principal stages, the ion exchange of molybdate ions (MoO$_4^{2-}$) with the Zeolitic Imidazolate Framework-L (ZIF−L) and the pyrolytic process of the precursor (as shown in Figure 1). Specifically, it starts with synthesizing ZIF−L wires on the surface of carbon cloth using a solvent method at room temperature. As shown in Figure 2a and Figure S1, high-density ZIF−L nanowires are distributed uniformly on the surface of the carbon cloth fibers. Subsequently, a fraction of the cobalt sites in ZIF−L are replaced by MoO$_4^{2-}$ to produce the Mo/ZIF−L structure through ion exchange. Specifically, in the weak base solutions, the 2-MI ligand in ZIF−L will bind to H$^+$ (due to reversible hydrolysis of H$_2$O), triggering the diffusion of a small amount of Co$^{2+}$ outwards.[18, 19] Afterwards, some cobalt vacancies are formed, and MoO$_4^{2-}$ will enter the cobalt vacancy and bond with unsaturated cobalt ions in the neighboring sites.[20] During this process, the morphology of Mo/ZIF−L is almost identical to that of ZIF−L, except that the diameter of the nanowires decreases (Figure S2a and S2b). This is due to the partial hydrolysis of the ZIF−L surface during the long-term ion exchange process, which is also proved by the TEM analysis (Figure S3). The shift of the XRD peaks to smaller angles signifies that some of the interplanar distances in ZIF−L have increased, indicating the successful incorporation of MoO$_4^{2-}$ into the lattice (as demonstrated in Figure S2c and S2d). Finally, at high temperature, the released N atoms from the 2-methylimidazole react with Mo and Co atoms to form Co$_3$Mo$_3$N and Co$_4$N and the C-skeleton of the organic ligands forms a carbon framework. In addition, some cobalt ions are reduced to metallic cobalt by carbon at high temperature. In Figure 2b and Figure S4, after high-temperature pyrolysis, the wires basically maintain their morphology, and it is reported that this wire array can prevent gas bubbles from accumulating on the catalyst surface, which facilitates mass diffusion.[21] Further, the crystallinity and phase purity of the pre-fabricated products are measured by XRD. In Figure 2c, all the samples show a peak at 26°, which can be attributed to the (002) plane of graphitic carbon (PDF#75-0444).[22] In Figure S5, for Co$_4$N/Co, the diffraction maxima at 35.5°, 44°, 52°, and 76° match well with the cubic phase of Co$_4$N with the space group *Pm3m* (ICDD 04–021-6262),[10] indicating its presence in the sample. However, the peaks at 44°, 52° and 76° are slightly shifted compared to Co$_4$N and are between the main peaks of cubic cobalt and Co$_4$N, indicating that Co$_4$N and Co can be present at the same time. For Co$_3$Mo$_3$N/Co$_4$N/Co, except for the graphitic carbon, Co, and Co$_4$N phases, all of the characteristic peaks are well compatible with the *FCC* Co$_3$Mo$_3$N phase with the space group *Fd3m* (PDF#89-7953).[23] Furthermore, no other obvious peaks were detected, demonstrating that the nanocomposites are composed of Co$_3$Mo$_3$N, Co$_4$N, Co, and graphite. Raman results also confirm the presence of graphite, Co$_4$N, and Co$_3$Mo$_3$N (Figure S6). In addition, Co$_3$Mo$_3$N/Co$_4$N/Co at different pyrolytic temperatures were also characterized by SEM, XRD and Raman spectroscopy (Figure S7 and Figure S8).

To further reveal detailed information about the interfaces in the metallic Co$_3$Mo$_3$N/Co$_4$N/Co heterostructure, cross-sectional transmission electron microscopy (TEM) was conducted. In Figure 2d, the bright-field TEM image of a single wire, as they are visible in the SEM image of Figure 2b, clearly shows several interfaces. One of them is enlarged in the high-resolution image of Figure 2e. Element distribution analysis based on energy-dispersive X-ray spectroscopy (EDXS) indicates interfacing Co and Co−Mo−N regions (Figure 2g–2j). It should be mentioned that the N signal in the crystalline Co region (Figure 2f) is at the noise level, so a potential N content is below the detection limit of about 0.1 at.%.

The composition of the synthesized samples was analyzed by X-ray photoelectron spectroscopy (XPS). In Figure S9a, a clear contribution of the Mo 3d peak is observed in the survey spectrum of the Co$_3$Mo$_3$N/Co$_4$N/Co heterostructure compared to Co$_4$N/Co. In Figure 3a, the high-resolution Co 2p spectrum for Co$_4$N/Co exhibits two peaks at binding energies of 779.7 eV and 794.8 eV, which were attributed to the metallic Co. In addition, the prominent peaks at 781 eV and 796.5 eV and the tiny peaks at 787.4 eV and 803.5 eV were ascribed to the Co−N bond and satellite peaks, respectively.[24] It is well known that $Co_3Mo_3N$ is composed of $Co_8$ clusters and $Mo_3N$, meaning that there are no Co−N bonds within its composition.[25] However, compared with $Co_4N$/Co, the Co−N peak of $Co_3Mo_3N$/$Co_4N$/Co shifts to a higher binding energy, suggesting that the introduction of $Co_3Mo_3N$ can lead to an electron redistribution at the $Co_3Mo_3N$/$Co_4N$ interface. Moreover, the high-resolution N 1s spectra show two peaks located at 398.5 and 396 eV, corresponding to Co−N and N−H bonds in $Co_4N$/Co (Figure 3b).[26] After introducing $Co_3Mo_3N$, two new peaks at 394 eV and 397.8 eV are observed, corresponding to the Mo 3p region and the Mo−N bond.[27] More importantly, the Co−N peak shifts towards a lower binding energy, which again proves that $Co_3Mo_3N$ and $Co_4N$ have a strong interaction. In addition, in the high-resolution Mo 3d spectrum of $Co_3Mo_3N$/$Co_4N$/Co (Figure 3c), the peaks at 228.1/231.5 eV, 229/232.4 eV, and 235.5 eV are assigned to $Mo^0$, $Mo^{\delta+}$, and satellite, respectively.[27] The carbon peak did not change after the introduction of $Co_3Mo_3N$, indicating that the carbon in the sample did not react with the metal ions (Figure S9b). In addition, the high-resolution C 1s spectrum exhibited three peaks at 284.30, 284.95, and 286.05 eV, demonstrating the presence of alternating C=C and C−C commensurate to graphene, C−O, and C=O, respectively.[28] The afore-mentioned XPS results confirm that the induced $Co_3Mo_3N$ can lead to a redistribution of the interface charges between $Co_3Mo_3N$ and $Co_4N$, which may be beneficial for HER and OER activity.

To further clarify the change of charges between $Co_3Mo_3N$ and $Co_4N$ in the heterostructure, ultraviolet photoemission spectroscopy (UPS) measurements were performed. As shown in Figure S10, the density of filled states of $Co_3Mo_3N$/$Co_4N$/Co and $Co_4N$/Co extends to the Fermi level, confirming that these materials have metallic properties.[29] Furthermore, the work function was calculated according to the equation of $W=h\nu–E_{cutoff}$, where $h\nu$ is the incident photon energy (21.22 eV) and $E_{cutoff}$ is the normalized secondary electron cutoff (Figure 3d). Values of 4.33 and 4.48 eV were determined for $Co_3Mo_3N$/$Co_4N$/Co and $Co_4N$/Co, respectively. In Figure S11, the work function of $Co_3Mo_3N$ is 4.09 eV, which is calculated by density functional theory

(DFT). Therefore, it is to be expected that electrons will be transferred from $Co_3Mo_3N$ to $Co_4N/Co$ due to the difference in work functions (Figure 3e). In addition, the Raman peak of $Co_4N$ in the $Co_3Mo_3N/Co_4N/Co$ heterostructure shifts to lower wavenumber compared to pure $Co_4N/Co$, which also shows that electrons are injected into $Co_4N$ (Figure S6a). It is reported that the Co sites in $Co_4N$ have a too strong adsorption force for H, which is not conducive to the desorption of $H_2$.[9] The flow of electrons into $Co_4N$ is beneficial to reduce the bond energy of the Co−H, thereby improving the HER efficiency.

The HER and OER performances of the various catalysts were evaluated in 1 M KOH with a standard three-electrode system. As shown in Figure 4a, $Co_3Mo_3N/Co_4N/Co$ displays small overpotentials of −78 mV for achieving current densities of −10 mA cm$^{-2}$ during HER, which is close to the value of the state-of-the-art Pt/C catalyst. The HER performance of $Co_3Mo_3N/Co_4N/Co$ is better than $Co_4N$, Co, $Co_4N/Co$, $Co_3Mo_3N/Co_4N$ and $Co_3Mo_3N/Co$ (Figure S12 and Figure S13), meaning that the synergy of $Co_3Mo_3N$, $Co_4N$ and Co can improve the HER activity. After adjusting the pyrolysis temperature of the Mo/ZIF−L precursor, $Co_3Mo_3N/Co_4N/Co$-750 shows the best HER performance. In addition, the Tafel slope of $Co_3Mo_3N/Co_4N/Co$-750 is 68 mV dec$^{-1}$, which is lower than that for $Co_3Mo_3N/Co_4N/Co$-700, $Co_3Mo_3N/Co_4N/Co$-800, and $Co_4N/Co$ and comparable with commercial Pt/C, indicating the fastest reaction kinetics of $Co_3Mo_3N/Co_4N/Co$-750 toward HER (Figure 4b). The electrochemically active surface area (ECSA) is a vital component for the HER activity. The ECSAs of the catalysts were determined based on the double-layer capacitances ($C_{dl}$) obtained through CV measurements (Figure S14a–d). Accordingly, the double-layer capacitance ($C_{dl}$) of $Co_3Mo_3N/Co_4N/Co$-750 is 15.4 mF cm$^{-1}$, and hence, 4.05, 2.08, and 1.37 times higher than that of $Co_4N/Co$, $Co_3Mo_3N/Co_4N/Co$-700, and $Co_3Mo_3N/Co_4N/Co$-800, respectively (**Figure S14e**). Therefore, the corresponding ECSA of $Co_3Mo_3N/Co_4N/Co$-750 is 256 cm$^2$ and thus higher than for the other catalysts, revealing more exposure of active sites (Figure S14f). Notably, $Co_3Mo_3N/Co_4N/Co$ still shows improved HER activities compared to $Co_4N/Co$, even after the current densities are ECSA-normalized (Figure S15), indicating its higher intrinsic activity. More importantly, the low overpotential of 78 mV in alkaline medium for $Co_3Mo_3N/Co_4N/Co$-750 outperforms most reported alkaline HER electrocatalysts (Table S1). To evaluate the structure stability of $Co_3Mo_3N/Co_4N$-750 during HER, SEM image and XRD analysis were conducted after a long-term stability test (Figure S16 and S17). These results show that $Co_3Mo_3N/Co_4N$-750 is characterized by an extremely high morphological and structural stability.

Interestingly, $Co_3Mo_3N/Co_4N/Co$ can also exhibit superior OER performance in alkaline media. According to Figure 4c and Figure S18, $Co_3Mo_3N/Co_4N/Co$-750 displays a low overpotential of 220 mV at 10 mA cm$^{-2}$, much smaller than that of commercial $RuO_2$ and the other catalysts of this study. Moreover, the Tafel slope is only 65 mV dec$^{-1}$ for $Co_3Mo_3N/Co_4N/Co$-750, i.e., much lower than that of commercial $RuO_2$ and $Co_4N/Co$, indicating superior reaction kinetics of $Co_3Mo_3N/Co_4N/Co$-750 for alkaline OER. Similar to the HER process, $Co_3Mo_3N/Co_4N/Co$ still shows improved OER activities compared to $Co_4N/Co$, even after the current densities are ECSA-normalized (Figure S19), indicating their higher OER intrinsic activities. Importantly, $Co_3Mo_3N/Co_4N/Co$-750 shows with 220 mV one of the lowest overpotential of compared to reported alkaline OER electrocatalysts (Table S1). In order to investigate the intrinsic catalytic activity of the electrocatalysts, the turnover frequencies (TOF) of HER and OER were evaluated at different potentials (Figure 4e and Figure S20 and S21). At an overpotential of −200 mV, the HER-TOF value of $Co_3Mo_3N/Co_4N/Co$-750 is around 4.6 $H_2$ s$^{-1}$, which is better than that of $Co_4N/Co$ (0.645 $H_2$s$^{-1}$). Besides, the OER-TOF value of $Co_3Mo_3N/Co_4N/Co$-750 (3.16 $O_2$ s$^{-1}$) is also higher than that of $Co_4N/Co$ (0.42 $O_2$ s$^{-1}$) at 400 mV. The above results indicate the excellent intrinsic catalytic activity of the $Co_3Mo_3N/Co_4N/Co$-750 electrocatalyst.

Motivated by its impressive performance, $Co_3Mo_3N/Co_4N/Co$-750 was utilized as both anode and cathode electrocatalyst in a two-electrode cell toward alkaline overall water splitting (Figure 4f). Commercial $RuO_2$ and Pt/C were used for comparison. Notably,

Co$_3$Mo$_3$N/Co$_4$N/Co-750||Co$_3$Mo$_3$N/Co$_4$N/Co-750 exhibits excellent activity toward overall water splitting. In 1 M KOH, the cell potential is 1.58 V at 10 mA cm$^{-2}$. To achieve the same current density, a much higher cell potential is needed for commercial RuO$_2$||Pt/C. In order to check whether all electrons are used in both the HER and OER processes during water splitting, we collected H$_2$ and O$_2$ separately using an H-type electrolytic cell and calculated their Faraday efficiencies (Figure S22). In Figure 4g, the Faraday efficiencies of both the HER and OER process are close to 100 %, proving that there are no side reactions in the processes. Furthermore, the electrochemical durability of Co$_3$Mo$_3$N/Co$_4$N/Co-750||Co$_3$Mo$_3$N/Co$_4$N/Co-750 and commercial RuO$_2$||Pt/C were explored by chronopotentiometry tests at different current densities in 1 M KOH. For commercial RuO$_2$||Pt/C, the cell potential strongly increases after 1 h (Figure S23), which is ascribed to the unstable crystal structure of RuO$_2$ and also consistent with literatures.[11, 30, 31] In contrast, Co$_3$Mo$_3$N/Co$_4$N/Co-750||Co$_3$Mo$_3$N/Co$_4$N/Co-750 can deliver superior durability over a wide range of current densities (0–200 mA cm$^{-2}$) and is stable up to 100 h, even at a current density of 200 mA cm$^{-2}$ (Figure 4h and 4i).

To gain deeper insight into the HER and OER mechanisms, operando Raman spectroscopy was utilized at different potentials. During the HER process, the in situ electrochemistry-Raman (EC-Raman) measurements were performed at the reduction potential to achieve surface structure information of Co$_3$Mo$_3$N/Co$_4$N/Co-750 and Co$_4$N/Co. It is reported that the protons of the alkaline HER originate from the water dissociation process.[32] Therefore, it is extremely important to understand the evolution of interfacial water molecules on the catalyst surface. First, in situ Raman spectroscopy testing of the support (carbon cloth) was used to rule out the influence of carbon cloth on the evolution of interfacial water (Figure S24). In Figure S24a, the two peaks located at 1350 and 1580 cm$^{-1}$ are associated with the D and G bands of carbon.[33] After applying the potential, no new peaks appear, indicating that the carbon cloth has no effect on the adsorption of H$_2$O during the HER process. Because the Raman peak of H$_2$O overlaps with the G band of carbon, D$_2$O was used to perform tests for in situ Raman experiments. In Figure 5a, when the catalyst is immersed into KOH solution, the vibrational peak of D$_2$O increases and shifts from 1200 cm$^{-1}$ to 1225 cm$^{-1}$ as the reduction potential increases, indicating the formation of asymmetric H$_2$O adsorption on the catalyst with H-down structure. In Figure 5b, the Raman spectra of Co$_3$Mo$_3$N/Co$_4$N/Co-750 and Co$_4$N/Co show a broad peak from 3200 cm$^{-1}$ to 3600 cm$^{-1}$ assigned to the O−H stretching mode of interfacial water. This interfacial water peak can be described with three peaks (Figure S25 and S26). In particular, the peaks at 3225 cm$^{-1}$, 3450 cm$^{-1}$, and 3560 cm$^{-1}$ belong to the tetrahedral, trihedral, and free water.[34] The free water has the H-down structure, indicating it is readily adsorbed on the catalyst surface. It is well known that the larger the proportion of free water, the lower the energy required for the dissociation of water, since the activation energy of free water is smaller than that of tetrahedrally and trihedrally coordinated water.[35] According to Figure 5c, for Co$_3$Mo$_3$N/Co$_4$N/Co-750, when the voltage was increased from −0.75 to −0.85 V, the proportion of free water (3560 cm$^{-1}$) also increased, indicating that it was easy to disrupt the hydrogen-bonding structure of tetra- and tri-coordinated water, thus generating more free water. Compared to Co$_4$N/Co, the faster growth rate of free water on the surface of Co$_3$Mo$_3$N/Co$_4$N/Co-750 indicates a faster water adsorption rate. When the voltage was increased to −0.9 V, the water dissociation process and the free water transfer process to the catalyst surface proceeded simultaneously. Therefore, the decrease in the free water peak is due to the faster rate of hydrolysis than the adsorption of water. However, for Co$_4$N/Co, the free water peak starts to decrease significantly at −1 V. The free water peak decreases at a faster rate for Co$_3$Mo$_3$N/Co$_4$N/Co-750 than for Co$_4$N/Co, indicating that it decomposes water faster. Accordingly, in Figure 5d, the protons adsorbed on the Co$_3$Mo$_3$N/Co$_4$N/Co-750 surface were detected, while the adsorbed protons were not detected on the Co$_4$N/Co surface due to its slow hydrolysis rate. Finally, the adsorbed H on the Co$_3$Mo$_3$N/Co$_4$N/Co-750 surface will combine with another water molecule to produce H$_2$.

Further analysis of the improved OER performance was done using in situ EC-Raman spectroscopy. The nitrides of the transition metals tend to form a thin layer of the corresponding (hydro)oxides at anodic currents when they get in contact with strong alkaline solutions, which means these new species are likely to play a vital role in the OER.[4] In Figure 5e and **5 f**, for $Co_3Mo_3N/Co_4N/Co$-750, the new phase β-CoOOH appeared when the potential was increased to 1.5 V. When the potential was increased to 1.7 V, partial conversion of β-CoOOH into γ-CoOOH was observed. For $Co_4N/Co$, only when the voltage was increased to 1.55 V, a weak β-CoOOH peak became visible. After the OER test, the CoOOH peak in both samples still existed, indicating the formation of an active phase on the catalyst surface (Figure S27). The XRD also proved the existence of CoOOH after the OER (Figure S28). Moreover, the Raman peak of β-CoOOH (508 cm$^{-1}$) in $Co_3Mo_3N/Co_4N/Co$-750 is shifted toward lower wavenumbers compared to that in $Co_4N/Co$ (511 cm$^{-1}$), indicating that the β-CoOOH phase is more likely to form in $Co_3Mo_3N/Co_4N/Co$-750. Therefore, the mixed β-CoOOH and γ-CoOOH phases that are more readily formed on the $Co_3Mo_3N/Co_4N/Co$-750 surface play a crucial role to the enhancement of the OER activity.

In order to gain more information on the OER kinetics at the electrocatalyst surface, in operando Electrochemical Impedance Spectroscopy (EIS) was performed. The equivalent circuit diagram of the two processes, i.e., the electrooxidation process (high frequency $10^5$–$10^1$ Hz) and the OER process (low frequency $10^1$–$10^{-1}$ Hz), is given in Figure S29. The solution resistance, the electrooxidation process resistance, and the OER process resistance are denoted as $R_s$, $R_1$, and $R_2$, respectively. The fitted Nyquist data according to the circuit diagram were plotted with increased potential from 1.2 to 1.6 V (Figure 5g and 5i). In Figure S30, $R_1$ for $Co_3Mo_3N/Co_4N/Co$-750 is smaller than for $Co_4N/Co$. This means that the introduction of $Co_3Mo_3N$ can lower the energy barrier of cobalt oxidation, which is also consistent with the operando Raman results. Moreover, the evolution of the reactant *OH on the catalyst surface can be described by $R_2$. In Figure 5i, $R_2$ for $Co_3Mo_3N/Co_4N/Co$-750 is much lower than for $Co_4N/Co$ in the whole potential range, revealing the faster kinetics for the adsorption of *OH. The pseudocapacitance arising from *OH is defined as $C_2$, which is utilized to quantify the adsorption coverage of *OH. $C_2$ of $Co_3Mo_3N/Co_4N/Co$-750 is higher than that of

Co$_4$N/Co in the whole potential range, indicating a higher *OH coverage for Co$_3$Mo$_3$N/Co$_4$N/Co-750. The rapid *OH accumulation of Co$_3$Mo$_3$N/Co$_4$N/Co-750 is expected to favor of overall catalytic driving force. Moreover, the adsorbed *OH on the electrocatalysts was evaluated based on the Laviron equation.[36] In Figure S31, the steady redox currents all exhibit a linear correlation with the square root of the scan rate in the CV. The k$_s$ value of Co$_3$Mo$_3$N/Co$_4$N/Co-750 is 0.097 s$^{-1}$, which is larger than that of Co$_4$N/Co, revealing the strong binding force of *OH. The above results suggest that following the incorporation of Co$_3$Mo$_3$N, Co$_3$Mo$_3$N/Co$_4$N/Co exhibits an enhanced ability to adsorb *OH, facilitating the formation of the genuine active oxide species CoOOH on its surface. Therefore, Co$_3$Mo$_3$N/Co$_4$N/Co-750 exhibits enhanced OER activity.

To further decipher the origin of Co$_3$Mo$_3$N/Co$_4$N/Co for promoting the HER and OER performance from a theoretical perspective, density functional theory (DFT) calculations were performed. According to the XPS analysis, the introduction of Co$_3$Mo$_3$N alters the electronic structure of Co$_4$N, implying that the Co$_3$Mo$_3$N/Co$_4$N heterointerface in Co$_3$Mo$_3$N/Co$_4$N/Co-750 is critical for the HER and OER enhancement. Therefore, Co$_4$N, Co$_3$Mo$_3$N, and Co$_3$Mo$_3$N/Co$_4$N models were constructed (Figure 6a and S32). Apparently, in Figure 6b, the coupling of Co$_4$N and Co$_3$Mo$_3$N leads to a local charge redistribution at the interface, which is consistent with the XPS and Raman analyses. By encouraging robust interfacial charge transfer, the *d*-band center ($\epsilon_d$) in the heterostructures can be adjusted effectively. The adsorption/desorption of protons as well as the water dissociation of active intermediates are both modulated by this adjustment. Therefore, the electrocatalytic performance is enhanced.

To reflect the change in $\epsilon_d$, the density of states (DOS) was examined. In Figure 6c, the electronic states of the Co$_3$Mo$_3$N/Co$_4$N heterostructure are greater in intensity than those of Co$_3$Mo$_3$N and Co$_4$N at the Fermi level (E$_f$), indicating an enhanced conductivity. Compared to Co$_3$Mo$_3$N and Co$_4$N, $\epsilon_d$ of Co$_3$Mo$_3$N/Co$_4$N shows an upshift closer to the Fermi level (Figure S33). The adsorption strength of the catalysts toward $H_2O$ reactants rises linearly with $d$-band center upshifts, in accordance with the $d$-band center theory.[37] Accordingly, $Co_3Mo_3N/Co_4N$ has the strongest adsorption strength toward $H_2O$ (Figure 6d). Due to the strong $H_2O$ adsorption on the $Co_3Mo_3N/Co_4N$ heterointerface, the H−O bond of the adsorbed $H_2O$ can be prolonged, facilitating the dissociation of $H_2O$ and significantly speeding up the slow Volmer step of alkaline HER. As expected, in Figure 6e, $Co_3Mo_3N/Co_4N$ exhibits the lowest water dissociation energy (0.9 eV), compared with $Co_3Mo_3N$ (3.6 eV) and $Co_4N$ (1.51 eV). For the alkaline HER, after the water dissociation process, the most important step is the adsorption of protons on the catalyst surface. In Figure 6f, the Gibbs free energy ($\Delta G_H$) of the N sites is closer to the thermoneutral value ($\Delta G_H=0$) than for other sites in the $Co_3Mo_3N/Co_4N$ heterostructure, meaning that the actually active sites are N sites not the Co and Mo sites. Also, $\Delta G_H$ of the N sites in the $Co_3Mo_3N/Co_4N$ heterostructure are closer to zero than that for $Co_3Mo_3N$ and $Co_4N$, suggesting the best HER performance. Then, the free energy of adsorption of oxygen intermediates was calculated based on different models to investigate the relationship between the electronic structure and OER performance. During the OER process, there are three intermediates (*OH, *O, and *OOH) formed. For the free energy diagram of $Co_3Mo_3N$ and $Co_4N$, the rate determining step (RDS) is the oxidation of *OOH to $O_2$. After introducing $Co_3Mo_3N$, the RDS for $Co_3Mo_3N/Co_4N/Co$ is also the oxidation of *OOH to $O_2$, but with a smaller $\Delta G$ of 1.61 eV. The decrease in $\Delta G$ indicates that the optimized electronic structure of heterogeneous interfaces is essential for improving the OER performance.

Based on the above analyses, the enhanced HER and OER mechanisms of the $Co_3Mo_3N/Co_4N/Co$ heterostructure can be summarized as shown in Figure 6h and **6 i**. First, the enhanced HER can be rationalized as follows. For $Co_4N/Co$, free water is first adsorbed at the cobalt sites to form Co-*$H_2O$. Further, the adsorbed water is decomposed to form Co−H and $OH^-$. The proton adsorbed at the Co site combines with another molecule of water and cleaves to form $H_2$. Although the Co sites in $Co_4N$ have a large $\Delta G_H$, the latter can be optimized by coupling with metallic cobalt. However, in this process, the Co site has a weak adsorption capacity for free water and requires high energy to cleave the water, resulting in a slow alkaline HER kinetics. When $Co_3Mo_3N$ is introduced, the heterogeneous interface formed by $Co_3Mo_3N$ and $Co_4N$ enhances the adsorption capacity of water, while decreasing the water dissociation energy. In addition, the optimized N atoms at the heterogeneous interface have suitable adsorption and desorption capacities for H, accelerating the rate of $H_2$ generation. As a result, $Co_3Mo_3N/Co_4N/Co$ exhibits excellent alkaline HER properties. Then, the enhanced OER can be summarized as follows. The introduction of $Co_3Mo_3N$ upshifts the *d*-band center of $Co_3Mo_3N/Co_4N/Co$ leading to increased adsorption of *OH. Meanwhile, the real active species β-CoOOH and γ-CoOOH are more easily formed compared to $Co_4N/Co$. As a result, $Co_3Mo_3N/Co_4N/Co$ exhibits excellent OER performance.

## 3 Conclusion

In order to accelerate the water electrolysis at high current densities, a metallic heterostructure $Co_3Mo_3N/Co_4N/Co$ was created with improved electron transfer efficiency and fast reaction dynamics. The upshift of the *d*-band center caused by the charge transfer in the $Co_3Mo_3N/Co_4N/Co$ heterostructure is advantageous for the adsorption of $H_2O$ and *OH. On the one hand, increased $H_2O$ adsorption on the surface of the heterostructure reduces the energy of water dissociation and accelerates proton production, hence enhancing the HER. The optimized interface electronic structure, on the other hand, makes it easier for the Co sites in the $Co_3Mo_3N/Co_4N/Co$ heterostructure to adsorb *OH species and produce the β-CoOOH and γ-CoOOH active phases, thus improving the OER performance. This work presents a new viewpoint on the construction of transition-metal-based metallic heterostructures for overall water splitting.

**Acknowledgments**

Y. Liu and L. Wang contributed equally to this work. The authors acknowledge Andreas Worbs for preparing the TEM lamella. Prof. Nikolai Gaponik is acknowledged for his valuable suggestions during the experiment. X. Zhang acknowledges the support from the National Natural Science Foundation of China (Grants No. 12274112). Y. Liu acknowledges the support from the China Scholarship Council (Grant No. 202106750023). Furthermore, the use of the HZDR Ion Beam Center TEM facilities and the funding of TEM Talos by the German Federal Ministry of Education and Research (BMBF, Grant No. 03SF0451) in the framework of HEMCP are acknowledged. Y. Vaynzof acknowledges the funding from the European Research Council (ERC) under the European Union's Horizon 2020 research and innovation programme (ERC Grant Agreement n° 714067, ENERGYMAPS). A. Eychmüller acknowledges support from the DFG (RTG 2767). Open Access funding enabled and organized by Projekt DEAL.

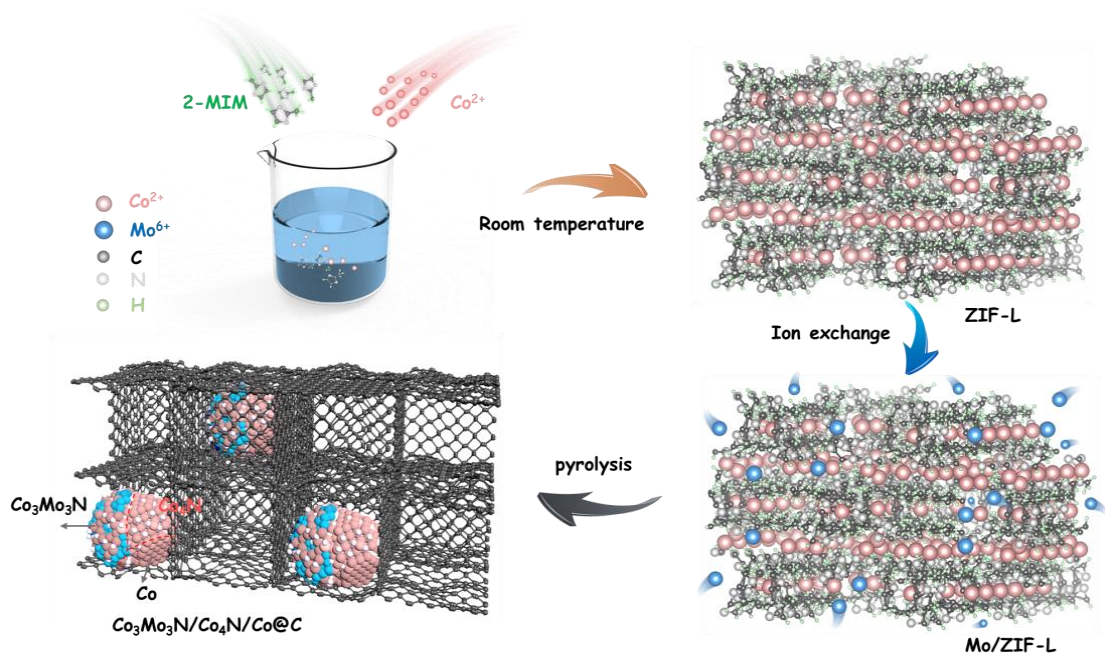

**Figure 1.** Schematic illustration of the synthetic process for the Co$_3$Mo$_3$N/Co$_4$N/Co.

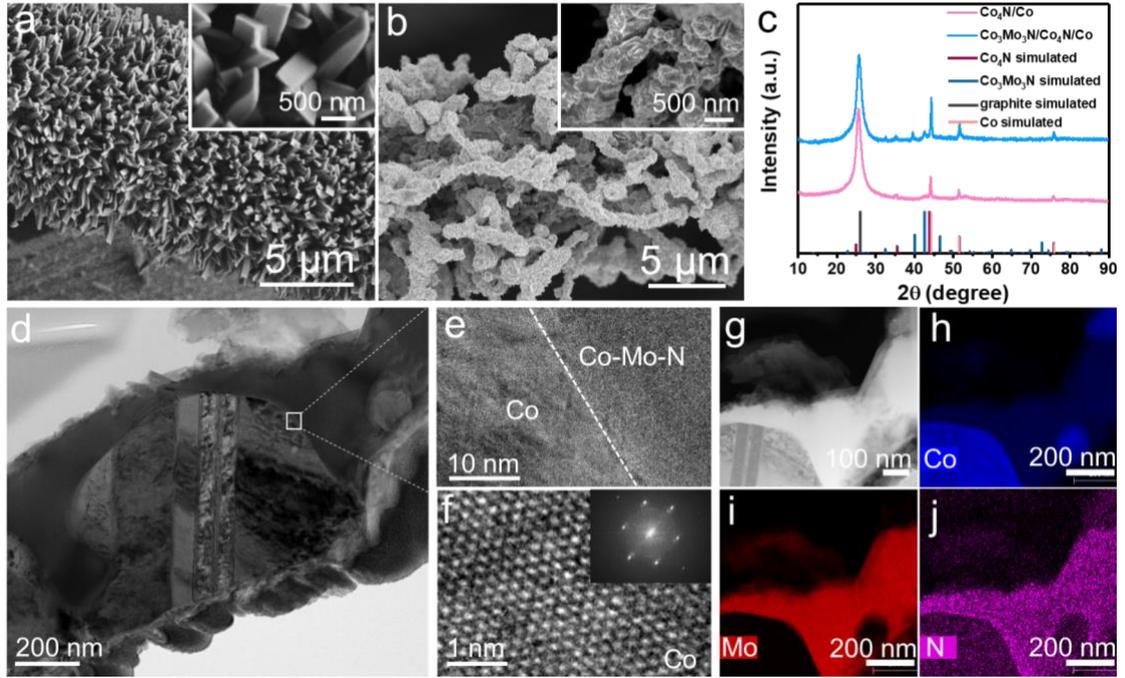

**Figure 2.** SEM image of (a) ZIF-L and (b) $Co_3Mo_3N/Co_4N/Co$ on carbon cloth. (c) XRD pattern of $Co_3Mo_3N/Co_4N/Co$ and $Co_4N/Co$. (d) Cross section of low magnification TEM image of $Co_3Mo_3N/Co_4N/Co$ wire; (e) interface region of Co and Co-Mo-N from d. (f) high-resolution TEM image from Co region. (g) FFT image from Co-Mo-N region. (h-k) EDS-mapping images of Co, Mo and N. Insert image in a and b are high magnification SEM images of $Co_3Mo_3N/Co_4N/Co$ and $Co_4N/Co$. Insert image in f is the FFT image from f.

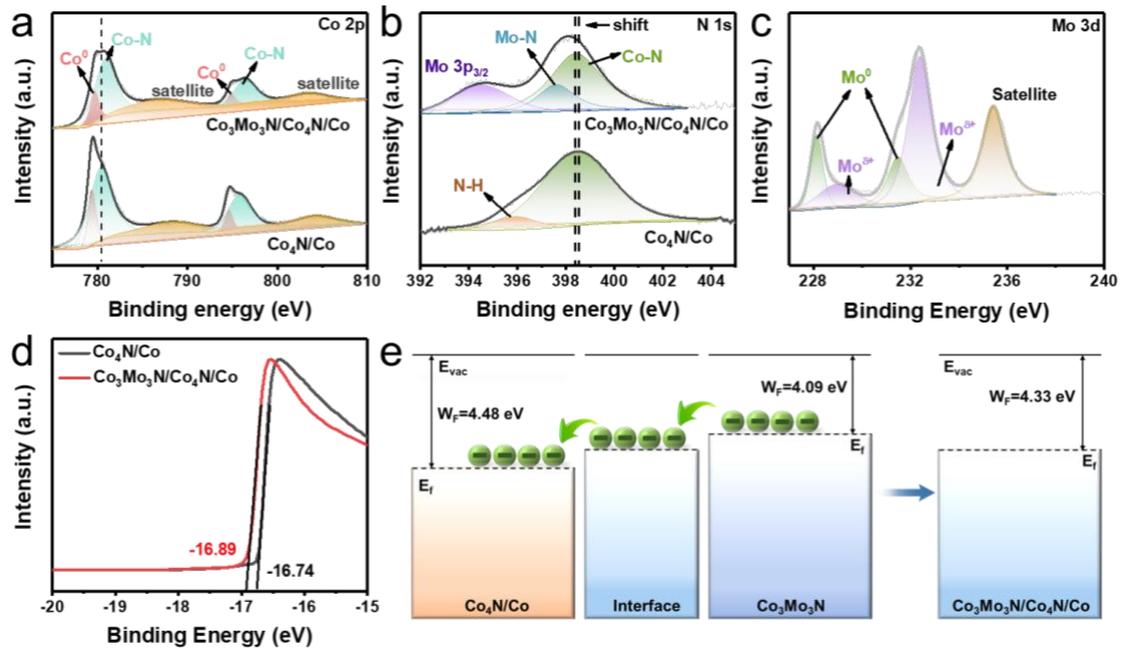

**Figure 3.** High resolution XPS pattern of (a) Co 2p, (b) N 1s and (c) Mo 3d in $Co_3Mo_3N/Co_4N/Co$ and $Co_4N/Co$. (d) UPS pattern of $Co_3Mo_3N/Co_4N/Co$ and $Co_4N/Co$. (e) Energy band diagrams of metallic $Co_4N/Co$ and $Co_3Mo_3N/Co_4N/Co$.

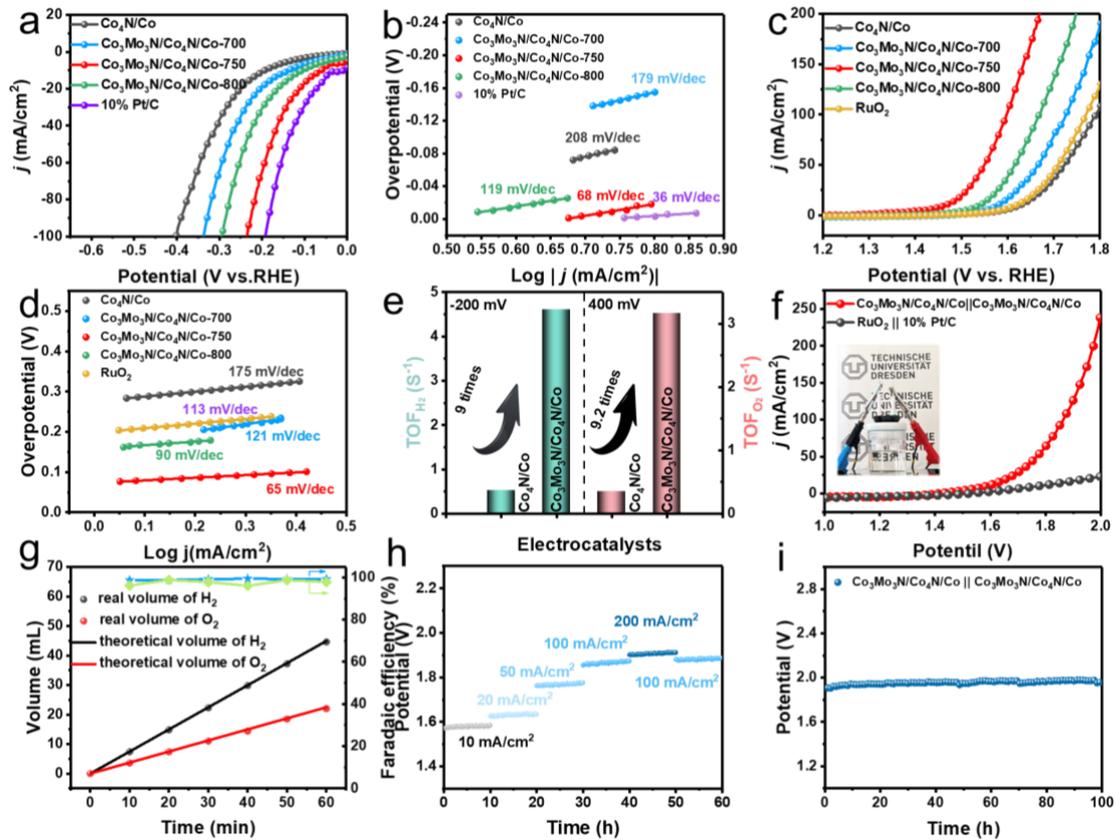

**Figure 4.** (a) HER polarization curves of Co$_4$N/Co, different pyrolysis time of Co$_3$Mo$_3$N/Co$_4$N/Co and 10% Pt/C. (b) Corresponding Tafel plots from a. (c) OER polarization curves of Co$_4$N/Co, different pyrolysis time of Co$_3$Mo$_3$N/Co$_4$N/Co and RuO$_2$. (d) Corresponding Tafel plots from c. (e) TOF value of HER and OER at -200 mV and 400 mV. (f) Overall water splitting performance of Co$_3$Mo$_3$N/Co$_4$N/Co || Co$_3$Mo$_3$N/Co$_4$N/Co and RuO$_2$ || 10% Pt/C with a two-electrode system. (g) Volume of gas produced in the HER and OER and corresponding Faraday efficiency. Blue and green lines represent the Faraday efficiency of H$_2$ and O$_2$, respectively; (h) Chronopotentiometry curves of Co$_3$Mo$_3$N/Co$_4$N/Co || Co$_3$Mo$_3$N/Co$_4$N/Co at different current densities; (i) Chronopotentiometry curves of Co$_3$Mo$_3$N/Co$_4$N/Co || Co$_3$Mo$_3$N/Co$_4$N/Co at 200 mA cm$^{-2}$.

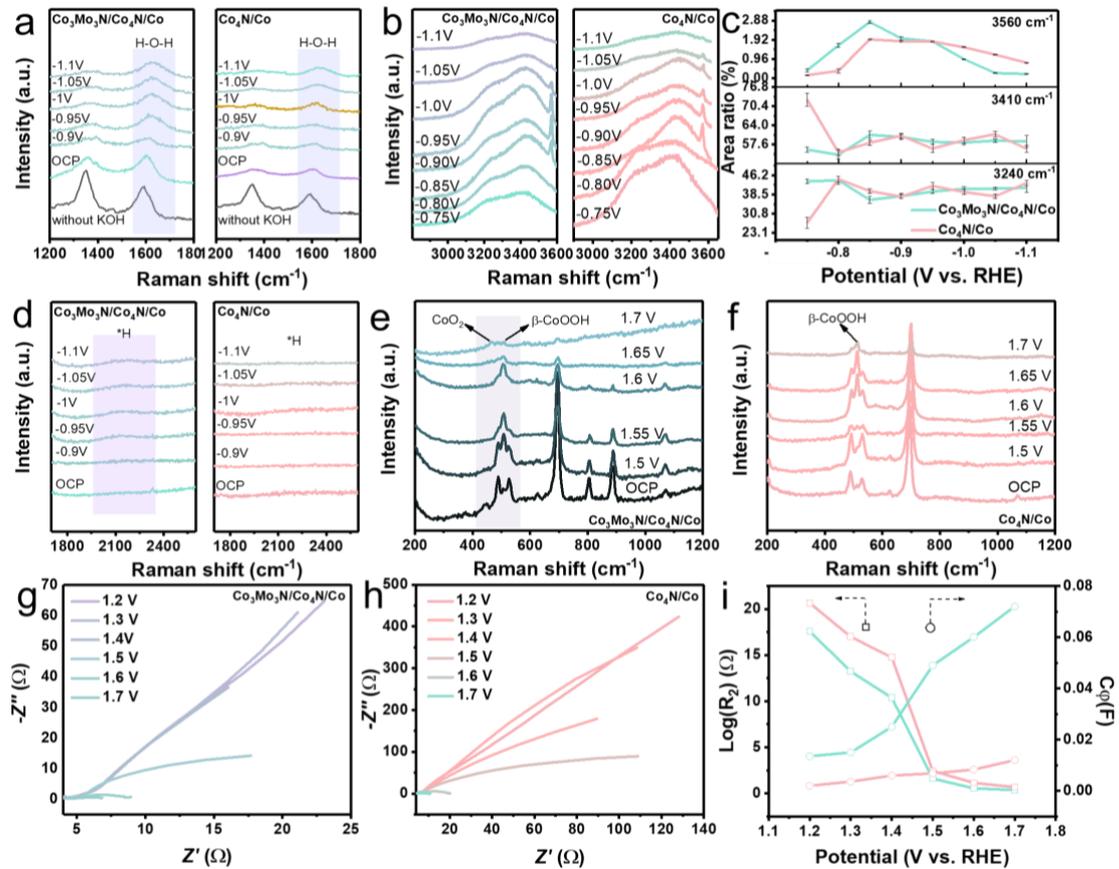

**Figure 5.** (a, b, d) The operando Raman spectra of $Co_3Mo_3N/Co_4N/Co$ and $Co_4N/Co$ at different potentials during HER. (c) Corresponding specific peak intensity of various peaks conducted from b. The operando Raman spectra of $Co_3Mo_3N/Co_4N/Co$ (e) and $Co_4N/Co$ (f) at different potentials during OER. Nyquist plots of $Co_3Mo_3N/Co_4N/Co$ (g) and $Co_4N/Co$ (h) at different potentials during OER; (i) the extracted $R_2$ and $C_φ$ from g and h.

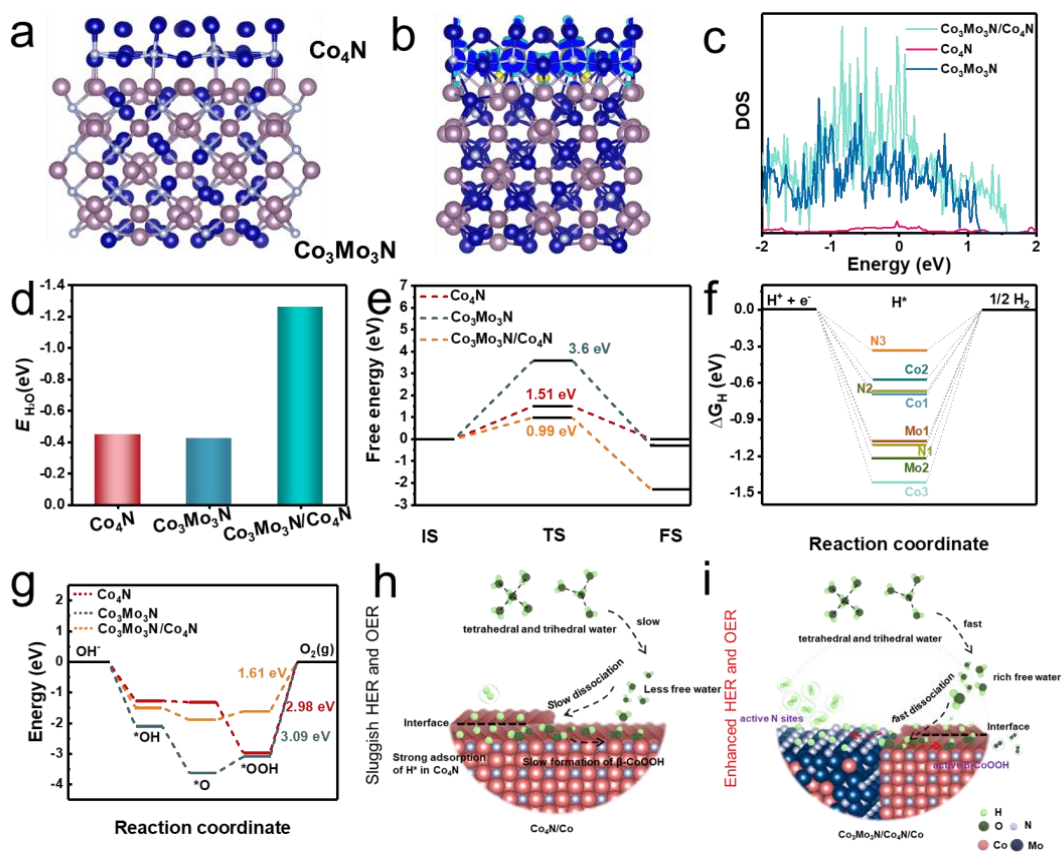

**Figure 6.** (a) The theoretical model of optimized $Co_3Mo_3N/Co_4N$. (b) Charge density difference plot at $Co_3Mo_3N/Co_4N$ heterointerface. Yellow bubbles for electron gain and blue bubbles for electron loss. (c) Density of state of $Co_3Mo_3N$, $Co_4N$ and $Co_3Mo_3N/Co_4N$. (d) water adsorption energy of $Co_3Mo_3N$, $Co_4N$ and $Co_3Mo_3N/Co_4N$. (e) water dissociation energy of $Co_3Mo_3N$, $Co_4N$ and $Co_3Mo_3N/Co_4N$. (f) Gibbs free energy of adsorbed H of different sites in $Co_3Mo_3N$, $Co_4N$ and $Co_3Mo_3N/Co_4N$. Co1, N1 are the sites in $Co_4N$. Mo1, Co2 and N2 are the sites in $Co_3Mo_3N$. Mo2, Co3 and N3 are the sites in $Co_3Mo_3N/Co_4N$. (g) Free energy diagram of OER for $Co_3Mo_3N$, $Co_4N$ and $Co_3Mo_3N/Co_4N$. (h) The sluggish OER mechanism of $Co_4N/Co$. (i) The enhanced OER mechanism of $Co_3Mo_3N/Co_4N/Co$.